\documentclass[aps,prb,preprintnumbers,amsmath,amssymb,superscriptaddress,twocolumn,10pt]{revtex4-2}
\usepackage{txfonts}
\usepackage{graphicx}
\usepackage{dcolumn}
\usepackage{bm}
\usepackage{array}
\usepackage[dvipsnames]{xcolor}
\usepackage[%
    colorlinks=true,
    pdfborder={0 0 0},
    linkcolor=blue,
    citecolor=blue,
    urlcolor=blue
]{hyperref}  
\usepackage{chngpage}
\usepackage{appendix}
\usepackage[caption=false]{subfig}
\usepackage{booktabs}

\newcommand{\expectation}[1]{\langle#1\rangle}

\newcommand{\be}{\begin{equation}}
\newcommand{\ee}{\end{equation}}
\newcommand{\bea}{\begin{eqnarray}}
\newcommand{\eea}{\end{eqnarray}}

\newcommand{\mbfk}{\mathbf{k}}

\begin{document}

\title{Emergence of nematic loop-current bond order in Kagome metals near van Hove singularities}

\author{Alex Friedlan}
\affiliation{Department of Physics, University of Toronto, 60 St. George Street, Toronto, Ontario, Canada, M5S 1A7}
\author{Hae-Young Kee}
\email{hy.kee@utoronto.ca}
\affiliation{Department of Physics, University of Toronto, 60 St. George Street, Toronto, Ontario, Canada, M5S 1A7}
\affiliation{Canadian Institute for Advanced Research, CIFAR Program in Quantum Materials, Toronto, Ontario, Canada, M5G 1M1}

\begin{abstract}
The recently-discovered family of Kagome metals has attracted significant interest due to reports of charge-bond order, orbital magnetism, and superconductivity. Some of these phases may exhibit time-reversal symmetry breaking. More recently, experiments have reported the emergence of nematic order that lowers the rotational symmetry of the system from sixfold to twofold. Here we investigate the mechanism behind a nematic loop-current bond order (NLCBO) that breaks both rotational and time-reversal symmetries. Examining an effective patch model that captures one $p$-type and one $m$-type van Hove singularity at each $M$ point, we find that frustration of the complex order-parameter phases leads to NLCBO. We further present conditions for overcoming other competing phases, including isotropic charge-bond and loop-current orders. Applying our findings to a previously studied model for $\mathrm{A}\mathrm{V}_3\mathrm{Sb}_5$ $(\mathrm{A}=\mathrm{K,Rb,Cs})$, we find that NLCBO emerges within a small region of phase space within mean-field theory. Our theory provides a microscopic description that goes beyond symmetry-allowed free-energy analyses and is broadly applicable to other Kagome metals featuring van Hove singularities near the Fermi level.
\end{abstract}

\date{\today}
\maketitle

\section{Introduction}

The Kagome lattice has long been celebrated as a platform for exploring geometric frustration, topology, and correlated electronic physics \cite{Wilson2024_AV3Sb5_Review,Guguchia2023_Kagome_CDW_SC,Wang2024_TopologicalKagome,KagomeReview2023_Intertwined}. At the most basic description, an idealized tight-binding model with a single $s$ or $d$ orbital at each Kagome site leads to topological features such as flat bands, Dirac crossings, and van Hove singularities  (vHSs) \cite{KimD2023}. While real materials necessarily deviate from this simple description, they may also exhibit modifications of these features that promote electronic correlations. 

The discovery of $\rm{A}\rm{V}_3\rm{Sb}_5$ \cite{Ortiz2019,WangReview2023} sparked an intense research program towards developing novel Kagome materials, and has since led to the identification of many new systems \cite{DiSante2025}. The $\rm{A}\rm{V}_3\rm{Sb}_5$ family of layered Kagome metals (alkali A = K, Rb, Cs) undergoes a phase transition to a $2a\times 2a$ charge-density wave (CDW), indicated by a kink in the in-plane resistivity on the order of $T_{\rm CDW}\sim 100$ K \cite{Ortiz2020}. In-plane and out-of-plane resistivity measurements differ by a factor of 600, indicating that these materials are quasi-two-dimensional, although the CDW may also order between layers \cite{Liang2021,Hu2022,JiangZ2023,Kato2022,XiaoQ2023}. The CDW is characterized by three ordering peaks $(3\mathbf{Q}),$ as reported in scanning tunneling microscopy (STM)  experiments \cite{Zhao2021}, and is not believed to possess long-range magnetic order \cite{Kenney2021}. As the temperature is lowered, these Kagome metals become superconducting at $T_c \sim 1$ K \cite{Jiang2021,Neupert2022,Wilson2024}. The superconducting state is likely unconventional \cite{Roppongi2023,Guguchia2023,Fukushima2024,Daniel2025,Nana2021,Deng2024,ChenH2021,ZhouS2022}, but a consensus has not yet been reached on either the symmetry of the gap or its microscopic origin.

Another notable case is the bilayer Kagome metal $\rm{Sc}\rm{V}_6\rm{Sn}_6,$ with CDW order that triples the unit cell, distinct from the $\rm{A}\rm{V}_3\rm{Sb}_5$ family \cite{Arachchige2022}. This material possesses multiple van Hove singularities near the Fermi level and exhibits nematic order \cite{Jiang2024,GuguchiaBilayer2023,Farhang2025}. Evidence for time-reversal-symmetry breaking (TRSB) in $\rm{Sc}\rm{V}_6\rm{Sn}_6$ remains controversial \cite{GuguchiaBilayer2023,Farhang2025}.
Another example is the magnetic Kagome metal $\rm{Fe}\rm{Ge},$ which also possesses multiple van Hove singularities near the Fermi level \cite{Teng2022,Teng2023}, and has opened the possibility of exploring the interplay of magnetism and charge ordering in Kagome materials.

Theoretically, various proposals for the origin of the CDW have been put forward, emphasizing electronic \cite{Kiesel2013} or phononic mechanisms \cite{Tan2021,Ferrari2022}. However, electronic order and lattice distortions likely go hand-in-hand. The CDWs in the Kagome metals have been associated with charge-bond order (CBO) \cite{Park2021,Christensen2021}, which can be understood as periodic modulations of the inter-site hopping amplitudes, and can be either real or complex. Real-valued modulations result in lattice distortions detectable by STM, and have been widely reported. More exotic is the so-called loop-current order (LCO), in which complex-valued modulations result in circulating orbital currents reminiscent of the Haldane model \cite{Haldane1988}, giving rise to TRSB lattice configurations with plaquettes carrying nonzero local flux. Some authors have reported TRSB in the CDW state that persists into the superconducting (SC) regime with $\mu$SR \cite{Yuarxiv2021,Mielke2022,Graham2024} and anomalous Hall \cite{YuF2021,Jiang2021NatureMat,TazaiPNAS2024} measurements. TRSB within the CDW state above $T_c$ was also reported in Magneto-optical Kerr effect (MOKE) measurements \cite{Wu2022PRB,XuY2022}, but remains controversial \cite{Saykin2023,Farhang2023, WangPRR2024}. See also \cite{Guo2022}.

Of primary interest for us are reports of electronic nematicity in the CDW state \cite{Xiang2021,Wu2022PRB,WuP2023}, lowering the rotational symmetry from $C_6$ to $C_2.$ Time-reversal symmetry may (CsV$_3$Sb$_5$ \cite{Nie2022}) or may not (KV$_3$Sb$_5$ \cite{Li2022}) be broken in the nematic state. Some authors have reported nematicity coinciding with the onset of CDW order \cite{Xiang2021}, but others have reported nematicity developing at lower temperatures ($T_{\rm nem}\sim 35$ K) \cite{Nie2022,Sur2023,ZhengL2022}. Meanwhile, Ref. \cite{Asaba2024} reported nematicity in the electronic state above the CDW transition based on magnetic torque measurements ($T_{\rm nem}\sim 130$ K). Nematic phases with both $2a\times 2a$ periodicity and $4a$ stripe order have been reported \cite{Li2023}. Interestingly, using laser-coupled STM, Ref. \cite{Xing2024} reported a nematic TRSB CDW whose chirality could be switched by an external electric or magnetic field. Within a Ginzburg-Landau framework, nematicity has been proposed to be a consequence of the simultaneous presence of LCO and CBO, referred to as loop-current bond order (LCBO), with the symmetry centres of each order out of phase \cite{TazaiNatComm2023}. Ref. \cite{Zhan2025} studied the ideal Kagome lattice for filling at the $p$-type vHS, including both nearest-neighbour (NN) and next-nearest-neighbour (NNN) Coulomb interactions within functional RG. The authors find CBO is favoured by NN interactions, while $\mathbf{Q}=0$ nematic charge order, $f$-wave superconductivity, and LCO with a large NNN component, develop as the NNN interaction strength is tuned. See also Refs. \cite{Feng2021,Feng2021PRB,Grandi2023,Grandi2024}.

In this paper we investigate the emergence of nematic loop-current bond order (NLCBO), characterized by the coexistence of nematic CBO and LCO within a single Kagome plane. Our central result is that NLCBO is stabilized by nearest-neighbour interactions within the effective patch model introduced by Li \textit{et al.} \cite{HeqiuPRL}, which captures the multiple vHSs near the $M$ points. In addition to NLCBO, we identify several other previously-reported competing orders, including the Star-of-David/Tri-Hexagonal CBO and LCBO. Our analysis of the effective model shows how nematic order can be stabilized by a strong anisotropic dispersion relative to the competing rotationally-symmetric phases, and highlights the frustration intrinsic to the Kagome lattice. To support our main findings, we also examine a tight-binding model for $\rm{A}\rm{V}_3\rm{Sb}_5,$ showing that the nematic phase persists once other bands are considered and the calculation is extended over the full Brillouin zone.

We assume that the ordering observed in the Kagome metals is $2a\times2a$ charge-bond order, arising from the nesting of $M$-point van Hove singularities. Accordingly, we consider only NN Coulomb interactions. This choice is motivated primarily for simplicity, as we find NN interactions alone to be sufficient for stabilizing a variety of CBO phases, including states that break time-reversal symmetry, rotational symmetry, or both. While further-neighbour hoppings have been argued to stabilize loop-current order \cite{Dong2023}, they do not appear to  be essential. We also neglect on-site interactions, as they necessarily act within sublattices of the same character and therefore cannot contribute to CBO. Moreover, on-site interactions are generally suppressed by the sublattice-interference effect \cite{Kiesel2012}.

The remainder of this paper is organized as follows. In Sec. \ref{effective2}, we introduce the effective model and the various CDW orders. In Sec. \ref{FSFX}, we present the phase diagram of the effective model and analyze the emergence of nematic order. In Sec. \ref{sec:9band}, we summarize the tight-binding model and present the mean-field phase diagram, showing that the nematic phase persists. In Sec. \ref{sec:Conclusion}, we discuss the implications of our results and outline directions for future research.

\section{Review of the Effective Patch Model and Order Parameters} \label{effective2}

\subsection{Effective patch model}
We examine the effective model introduced by Li \textit{et al.} \cite{HeqiuPRL} on patches near the three $M$ points in the Brillouin zone. The minimal kinetic contribution to the patch model that includes vH1, vH2, and their mixing $\lambda$, is given by
\begin{align}
    \mathcal{H}_{\rm patch} = & \sum_{\alpha \mbfk} \epsilon\left( \psi_{2\alpha\mbfk}^\dagger \psi_{2\alpha\mbfk} - \psi_{1\alpha\mbfk}^\dagger \psi_{1\alpha\mbfk} \right) \nonumber \\ + &\lambda k_\alpha \left(\psi_{1\alpha\mbfk}^\dagger \psi_{2\alpha\mbfk} + \psi_{2\alpha\mbfk}^\dagger \psi_{1\alpha\mbfk}\right),
\end{align}
where $\psi_{n\alpha\mbfk}^\dagger$ creates an electron at vH$n$ with momentum $\mbfk+\mathbf{M}_\alpha.$ The $k_\alpha$ are given explicitly by $k_1=-k_x/2+\sqrt{3}k_y/2,$ $k_2=-k_x/2-\sqrt{3}k_y/2,$ and $k_3=k_x,$ and are related by threefold rotation. The parameter $\lambda$ controls the degree of level repulsion between vH1 and vH2 and must have dimensions of energy $\times$ length. vH1 is a $p$-type vHS while vH2 is of $m$-type.
    
\begin{figure*}
    \centering
    \includegraphics[width=0.8\linewidth]{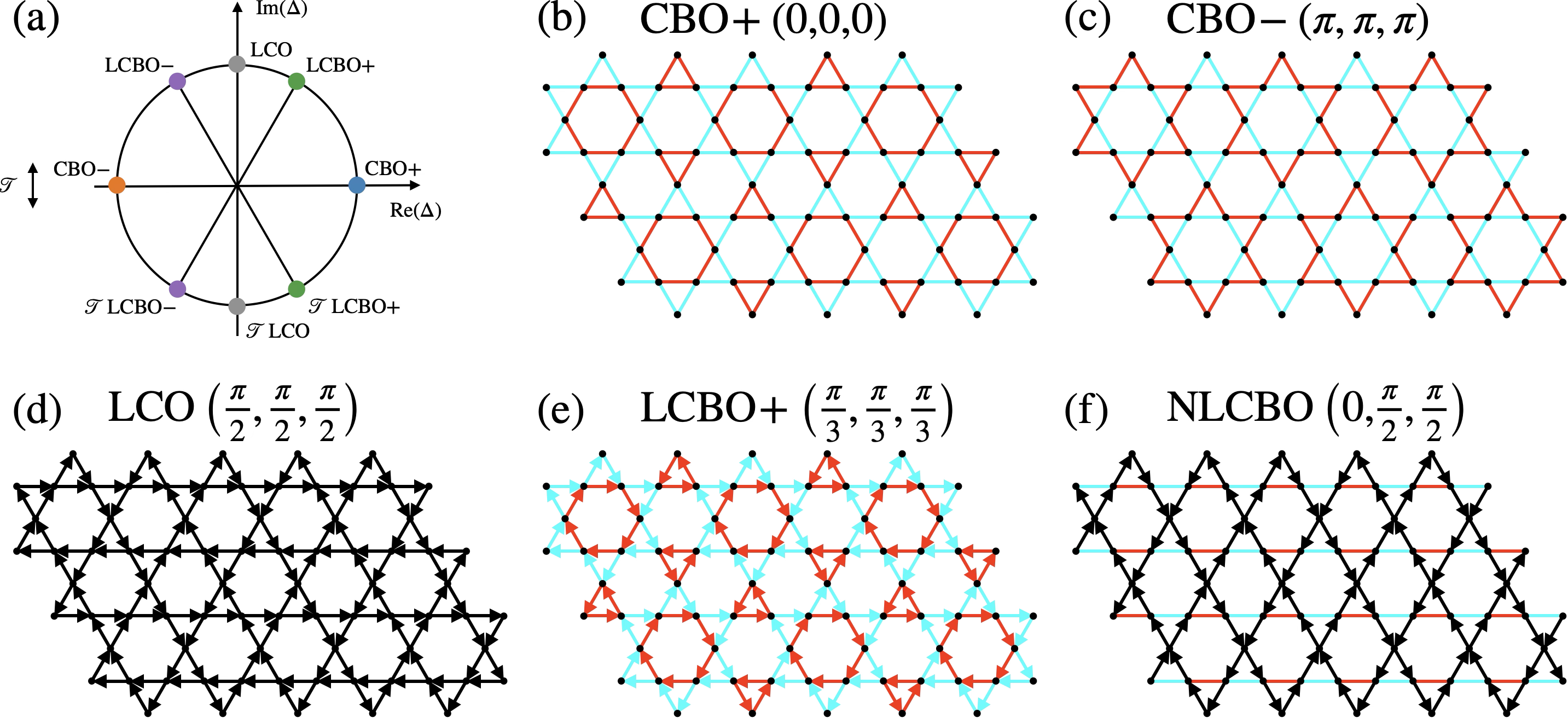}
    \caption{Summary of bond-ordered phases. (a) For uniform phases preserving rotational symmetry, the order parameter can be represented on the complex plane. Real-valued $\Delta$ corresponds to modulations of the hopping amplitude (red positive, blue negative): (b) $\Delta>0$, (c) $\Delta<0$. Complex-valued $\Delta$ have non-zero expectation value of the current density $j_{\alpha\beta}\sim i\expectation{c_{\mathbf{R}\alpha}^\dagger c_{\mathbf{R'}\beta}-c_{\mathbf{R'}\beta}^\dagger c_{\mathbf{R}\alpha}}.$ The direction of the current is represented by the arrows, and the TR partner is obtained by reversing the directions of the arrows. (d) Pure imaginary-valued $\Delta$ corresponds to a pattern of orbital currents without modulation of the hopping magnitude. (e) Coexisting loop-current and charge-bond order, possessing both amplitude and phase modulations. (f) Nematic loop-current bond order (ordering wavevector $\mathbf{Q}_{AB}=\mathbf{M}_C$ is chosen for example). The phase structure $(\phi_1,\phi_2,\phi_3)$ of each order is indicated (see text).}
    \label{fig_NLCBO}
\end{figure*}

The charge-bond order parameter is written \cite{Denner2022}
\be
    \Delta_{\alpha\beta}=\frac{V}{2N}\sum_{\mathbf R}\left(\langle d^\dagger_{\mathbf R,\alpha}d_{\mathbf R,\beta} \rangle -\langle d^\dagger_{\mathbf R,\alpha}d_{\mathbf R-\mathbf d_{\alpha\beta},\beta} \rangle \right)\cos(\mathbf Q_{\alpha\beta} \cdot \mathbf R),
\label{Deldef}
\ee
where $d_{\mathbf{R},\alpha}$ denotes a Kagome $d$ orbital on sublattice $\alpha$ in unit cell $\mathbf{R},$ and $\mathbf Q_{\alpha\beta}=\mathbf{M}_\alpha-\mathbf{M}_\beta$ connects the three $M$ points. The ordering described by Eq. (\ref{Deldef}) can be derived most naturally from a NN Coulomb interaction ${\mathcal{H}_V = V\sum_{\langle ij\rangle}\sum_{\alpha\neq\beta}n_i^\alpha n_j^\beta}$, which necessarily acts between Kagome sites of different sublattice character. The CDW phase is characterized by the triplet $(\Delta_{AB},\Delta_{BC},\Delta_{CA}).$ Projecting the interaction onto vH1 and vH2, one obtains
\begin{equation}
    \mathcal{H}_{\rm CDW} = \sum_{\mbfk,\alpha\neq\beta}  \left(s_1\Delta_{\alpha\beta}\psi_{1\alpha\mbfk}^\dagger \psi_{1\beta\mbfk} + s_2\Delta_{\alpha\beta}^*\psi_{2\alpha\mbfk}^\dagger \psi_{2\beta\mbfk}\right).
\end{equation}
The parameters $s_1=-2|b'|^2$ and $s_2=2|b|^2$ originate from the projection and are a direct consequence of the $p$- and $m$-type character of vH1 and vH2. Namely, symmetry considerations enforce the orbital composition of vH1 and vH2 to take the form (e.g. at $M_A$): $\psi_{1A\mbfk}=b'd_{\mbfk,A}+...$ and $\psi_{2A\mbfk}=b(d_{\mbfk,B}-d_{\mbfk,C})+...,$ where the ellipses denote a small admixture of ligand orbitals.

Altogether, the effective Hamiltonian is given by $\mathcal{H}_{\rm eff} = \mathcal{H}_{\rm patch}+\mathcal{H}_{\rm CDW}=\sum_\mbfk \Psi_\mbfk^\dagger \mathcal{H}(\mbfk)\Psi_\mbfk.$ In the basis $\left\{\psi_{2A\mbfk},\psi_{2B\mbfk},\psi_{2C\mbfk},\psi_{1A\mbfk},\psi_{1B\mbfk},\psi_{1C\mbfk}\right\},$ we have explicitly
\begin{equation}\label{explicit}
\mathcal{H}(\mbfk)=
    \begin{pmatrix}
        \epsilon & s_2\Delta_{AB}^* & s_2\Delta_{CA} & \lambda k_1 & 0 & 0 \\
        s_2\Delta_{AB} & \epsilon & s_2\Delta_{BC}^* & 0 & \lambda k_2 & 0 \\
        s_2\Delta_{CA}^* & s_2\Delta_{BC} & \epsilon & 0 & 0 & \lambda k_3 \\
        \lambda k_1 & 0 & 0 & -\epsilon & s_1\Delta_{AB} & s_1\Delta_{CA}^* \\
        0 & \lambda k_2 & 0 & s_1\Delta_{AB}^* & -\epsilon & s_1\Delta_{BC} \\
        0 & 0 & \lambda k_3 & s_1\Delta_{CA} & s_1\Delta_{BC}^* & -\epsilon \\
    \end{pmatrix}.
\end{equation}
In Eq. (\ref{explicit}), vH1 and vH2 are distinguished by the factors $s_1$ and $s_2,$ the energy separation $2\epsilon,$ and the relative complex conjugation of $\Delta.$ All three of these features arise directly from the geometry of the vanadium Kagome lattice, however their precise values are influenced by the presence of the ligand orbitals. While the relative sign between $s_1$ and $s_2$ is constrained by symmetry, their precise values are material-dependent. For example, $s_1$ and $s_2$ will depend on which member of the $\rm{A}\rm{V}_3\rm{Sb}_5$ family is being considered \cite{Sim2024}. In the present study, the tight-binding model gives $b'\approx 0.9$ and $b\approx0.5$ for $\rm{Cs}\rm{V}_3\rm{Sb}_5,$ implying that the antimony orbitals contribute significantly to vH2 but only slightly to vH1. Upon projecting the NN interaction among vanadium orbitals onto vH1 and vH2, the low-energy Hamiltonian inherits the orbital weight through $s_1=-2|b'|^2$ and $s_2=2|b|^2.$

\subsection{Summary of Ordered Phases}
For phases preserving rotational symmetry, the three CDW components are equal, with $\Delta_{AB}=\Delta_{BC}=\Delta_{CA}\equiv\Delta.$ However, $\Delta$ is a complex number and the (rotationally-symmetric) orders can be classified according to their complex phase, see Fig. \ref{fig_NLCBO}(a). The simplest case is the symmetric charge-bond order (CBO), with $\Delta\in\mathbb{R}.$ Physically, this CDW corresponds to a periodic modulation of the NN hopping amplitudes. Depending on the sign of $\Delta,$ the resulting pattern is called Star-of-David (CBO$-$) or inverse Star-of-David (CBO$+$), see Fig. \ref{fig_NLCBO}(b)-(c). CBO$+$ has a positive $\Delta$ on triangles and hexagons, denoted by red bonds in Fig. \ref{fig_NLCBO}(b), while CBO$-$ has positive $\Delta$ on the Stars-of-David, Fig. \ref{fig_NLCBO}(c). CBO$\pm$ preserve sixfold rotation, inversion, and time-reversal symmetries.

Another possibility is the loop-current order (LCO) phase with purely imaginary $\Delta,$ i.e. $\Delta=i|\Delta|.$ This CDW has the same periodicity as CBO, but introduces a complex Peierls phase to the hopping amplitudes, resulting in a pattern of circulating orbital currents, Fig. \ref{fig_NLCBO}(d). LCO also preserves sixfold rotational symmetry, but breaks time-reversal symmetry. The purely imaginary LCO phase does not stabilize within our mean-field calculations, but instead always appears in superposition with CBO \cite{Vanderbilt2025}. 

This third possibility of coexisting CBO and LCO, denoted LCBO, is characterized by $\Delta=|\Delta|e^{i\phi},$ see Fig. \ref{fig_NLCBO}(e). Within our calculations, the complex phase is almost always found to be $\tfrac{\pi}{3}$ or $\tfrac{2\pi}{3}.$ The real part of $\Delta$ gives a finite CBO, whose sign distinguishes CBO$\pm.$ The imaginary part yields orbital currents, whose sign distinguishes time-reversal partners. We classify LCBO according to the sign of its real component: LCBO$+$ has positive $\rm{Re}(\Delta)$ on triangles and hexagons, similar to CBO$+,$ while LCBO$-$ has positive $\rm{Re}(\Delta)$ on the Stars-of-David, similar to CBO$-$.

Finally, the nematic CDW is characterized by an order parameter of the form $(\Delta_{AB},\Delta_{BC},\Delta_{CA})=(\Delta,i\Delta',i\Delta'),$ with $\Delta,\Delta'\in \mathbb{R}.$ This nematic phase, denoted NLCBO, has coexisting LCO and CBO as shown in Fig. \ref{fig_NLCBO}(f). NLCBO breaks two of three mirror planes, reducing the rotational symmetry from $C_6$ to $C_2.$ Various other possibilities for $(\Delta_{AB},\Delta_{BC},\Delta_{CA})$ are allowed by symmetry \cite{Feng2021PRB}, but do not appear as free-energy minima within our calculations. The identification of this particular phase structure is justified in Sec. \ref{phaseD9x9}, and has been predicted previously under the name $2\mathbf{Q}-1\mathbf{Q}$ order \cite{Christensen2022}.

A useful quantity for characterizing the various orders is the total phase $\Phi\equiv \phi_1+\phi_2+\phi_3$ (mod $2\pi$), where the $\phi_i$ correspond to the complex phases of the order parameters, i.e. $\Delta_{AB}=|\Delta_{AB}|e^{i\phi_1},$ $\Delta_{BC}=|\Delta_{BC}|e^{i\phi_2},$ $\Delta_{CA}=|\Delta_{CA}|e^{i\phi_3}.$ In all ordered states obtained in this paper, $\Phi=0$ or $\pi.$ The $\Phi=0$ phases are CBO$+$ and LCBO$-,$ while the $\Phi=\pi$ phases are CBO$-,$ LCBO$+,$ and NLCBO. We emphasize that $\Phi$ does \textit{not} serve as an indicator of TRSB, which is instead encoded in a non-vanishing expectation value of the current-density operator through the individual $\phi_i.$ In particular, for the real-ordered phases (CBO$\pm$), $\Phi$ simply encodes the sign of the gauge-invariant product $\Delta_{AB} \Delta_{BC} \Delta_{CA},$ and does not imply TRSB.

\section{Frustration and mechanism for nematic order}\label{FSFX}

\subsection{Phase diagram of the effective model} \label{minimalmodelsec}

\begin{figure*}
    \centering
    \includegraphics[width=\textwidth]{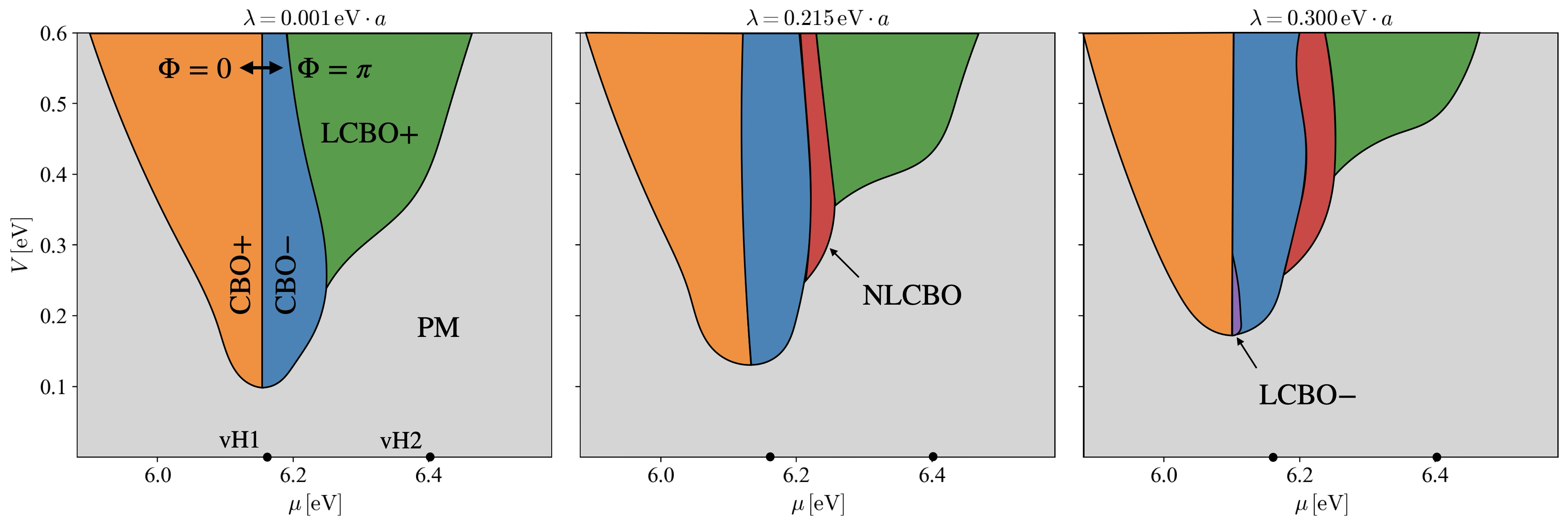}
    \caption{Phase diagram of the minimal model for three choices of $\lambda$ at $T = 90$ K. See Fig. \ref{fig_NLCBO} for a description of the ordered phases. PM refers to a pristine metal phase (disordered). The locations of the van Hove singularities are indicated.}
    \label{minimalmodel}
\end{figure*}

The phase diagram of the effective model is shown in Fig. \ref{minimalmodel} for three choices of the mixing $\lambda.$ Generally speaking, the phase diagram can be separated into two distinct regions based on the total phase $\Phi.$ As we elaborate on in Appendix \ref{Phitransition}, the total phase $\Phi$ transitions from $0$ to $\pi$ as the chemical potential passes through vH1. Within the $\Phi=0$ region, only CBO$+$ is stabilized. Within the $\Phi=\pi$ region, several competing phases develop (CBO$-$, LCBO$+$, and NLCBO), depending on $\mu, V,$ and $\lambda.$ In particular, NLCBO emerges in a small but extended window in between vH1 and vH2. We see that NLCBO emerges only when $\lambda$ is sufficiently large, indicating that the curvature of the bands near the $M$ points plays a crucial role in stabilizing the nematic order.

\begin{figure}
    \centering
    \includegraphics[width=\linewidth]{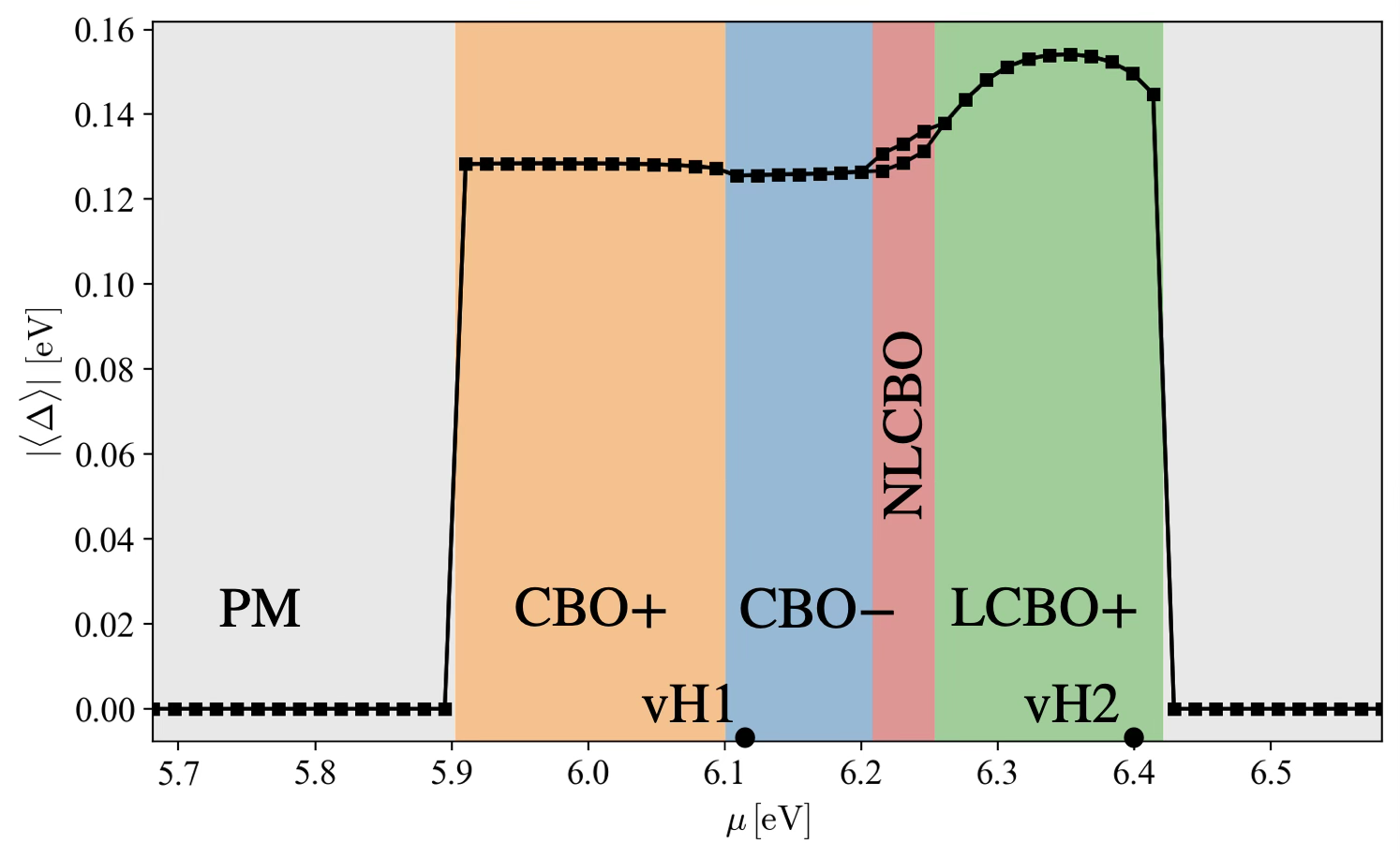}
    \caption{Order parameter amplitude along $V=0.5$ eV, $T = 90$ K, and $\lambda k_{\rm cut} = 0.3\, {\rm{eV}}\cdot a.$ For NLCBO, $\Delta'>\Delta.$}
    \label{fig:AmplitudeCut}
\end{figure}

In Fig. \ref{fig:AmplitudeCut}, we plot the amplitude of the order parameter along constant $V=0.5$ eV for $\lambda=0.3\,{\rm{eV\cdot}}a$, where $a$ is the lattice constant. There is a first-order transition from the pristine metal state (disordered) to the ordered state. The transitions between the various CDW phases appear to be second order. For the nematic phase, we see that the order-parameter amplitudes of the coexisting LCO and CBO are inequivalent ($\Delta\neq\Delta'$). We note that it is the phase structure, not the asymmetry of $\Delta$ and $\Delta',$ that is the primary force in stabilizing NLCBO. Indeed, if we fix $\Delta=\Delta'$ by hand, this equal-amplitude NLCBO phase still attains a lower free energy than the competing phases. Mean-field calculations are performed using a simulated annealing algorithm on a hexagonal $k$-space grid with approximately 500 points. We found this sampling sufficiently dense to obtain convergence. For more details, see Appendix \ref{Numerical}.

\subsection{Phase frustration as a mechanism for nematic order}

\begin{figure}
    \centering
    \includegraphics[width=0.9\linewidth]{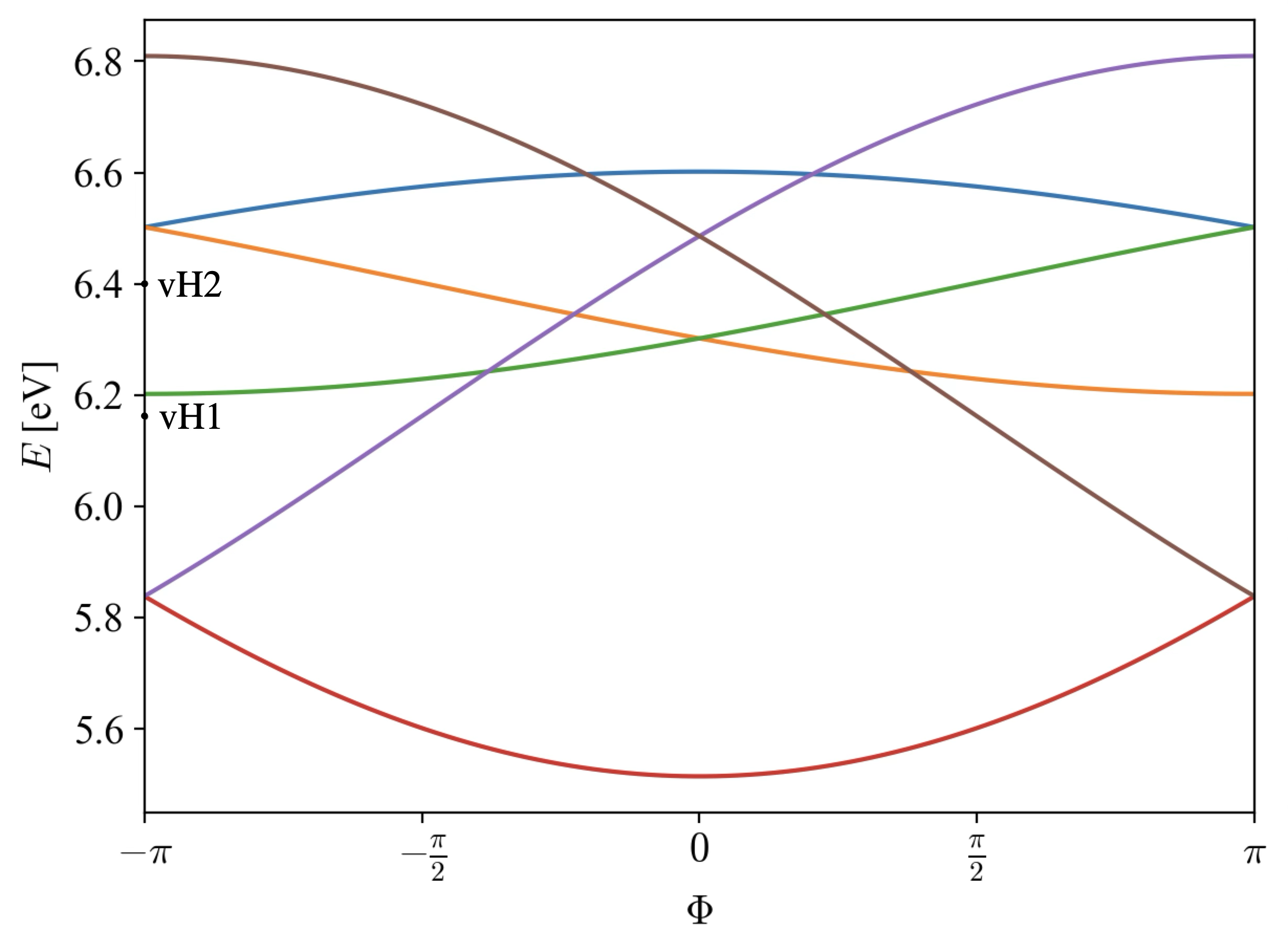}
    \caption{Eigenvalue spectrum of the unperturbed Hamiltonian as a function of the total phase $\Phi=\phi_1+\phi_2+\phi_3.$ The spectrum has degeneracies at $\Phi=0$ and $\Phi=\pi.$ Here we set $\Delta=0.2$ eV.}
    \label{fig:eigenvaluespecturm}
\end{figure}

The tendency to develop nematic order can be analyzed by the parameterization $\left(\Delta_{AB},\Delta_{BC},\Delta_{CA}\right)=\Delta (e^{i\phi_1},e^{i\phi_2},e^{i\phi_3}),$ where $\Delta$ is real and positive. We assume uniform $\Delta$ in this analysis to isolate the effect of phase frustration among the $\phi_i$. Treating $\lambda$ as a perturbation, we write Eq. (\ref{explicit}) as $\mathcal{H}(\mbfk)=\mathcal{H}_0+\lambda V(\mbfk),$ and perform a gauge transformation to express the unperturbed Hamiltonian $\mathcal{H}_0$ in terms of the total phase $\Phi=\phi_1+\phi_2+\phi_3.$ The transformation matrix is $U=\rm{diag}\left(1,e^{i\phi_1},e^{i(\phi_1+\phi_2)},1,e^{-i\phi_1},e^{-i(\phi_1+\phi_2)} \right),$ which yields
\begin{equation}
    U^\dagger \mathcal{H}(\mbfk)U = 
    \begin{pmatrix}
        \mathcal{H}_2 & \tilde{V}^\dagger(\mbfk) \\
        \tilde{V}(\mbfk) & \mathcal{H}_1
    \end{pmatrix},
\end{equation}
where
\begin{equation}
    \mathcal{H}_2=
    \begin{pmatrix}
        \epsilon & s_2\Delta & s_2\Delta e^{i\Phi} \\
        s_2\Delta & \epsilon & s_2\Delta \\
        s_2\Delta e^{-i\Phi} & s_2\Delta & \epsilon
    \end{pmatrix},
\end{equation}
\begin{equation}
    \mathcal{H}_1=
    \begin{pmatrix}
        -\epsilon & s_1\Delta & s_1\Delta e^{-i\Phi} \\
        s_1\Delta & -\epsilon & s_1\Delta \\
        s_1\Delta e^{i\Phi} & s_1\Delta & -\epsilon
    \end{pmatrix},
\end{equation}
and
\begin{equation}
    \tilde{V}(\mbfk)=
    \begin{pmatrix}
        k_1 & 0 & 0 \\
        0 & k_2 e^{2i\phi_1} & 0 \\
        0 & 0 & k_3 e^{2i(\phi_1+\phi_2)}
    \end{pmatrix}.
\end{equation}
At the cost of moving the $\phi_i$-dependence onto the perturbation, the transformation allows the $3\times3$ blocks associated with vH1 and vH2 to be expressed solely in terms of $\Phi.$ When $\lambda=0,$ the unperturbed eigenvalues of $\mathcal{H}_0$ are
\begin{equation}
    E_n^{(i)} = (-1)^i\epsilon -\mu + 2s_i\Delta\cos\left(\frac{1}{3}\left(\Phi+2\pi n\right)\right),
\end{equation}
where $n=0,1,2,$ labels the particular level originating from vH$i,$ with $i=1,2.$ This eigenvalue spectrum is shown in Fig. \ref{fig:eigenvaluespecturm} as a function of $\Phi$. One can show that the free energy of the unperturbed Hamiltonian is minimized at either $\Phi=0$ or $\Phi=\pi,$ depending on the chemical potential and $\Delta.$ In all cases, however, the total phase transitions from $\Phi=0$ to $\Phi=\pi$ as the chemical potential moves through vH1. This explains the sharp transition from CBO$+$ ($\Phi=0$) to CBO$-$ ($\Phi=\pi$) in Fig. \ref{minimalmodel} (see Appendix \ref{Phitransition}).

\begin{figure}
    \centering
    \includegraphics[width=\linewidth]{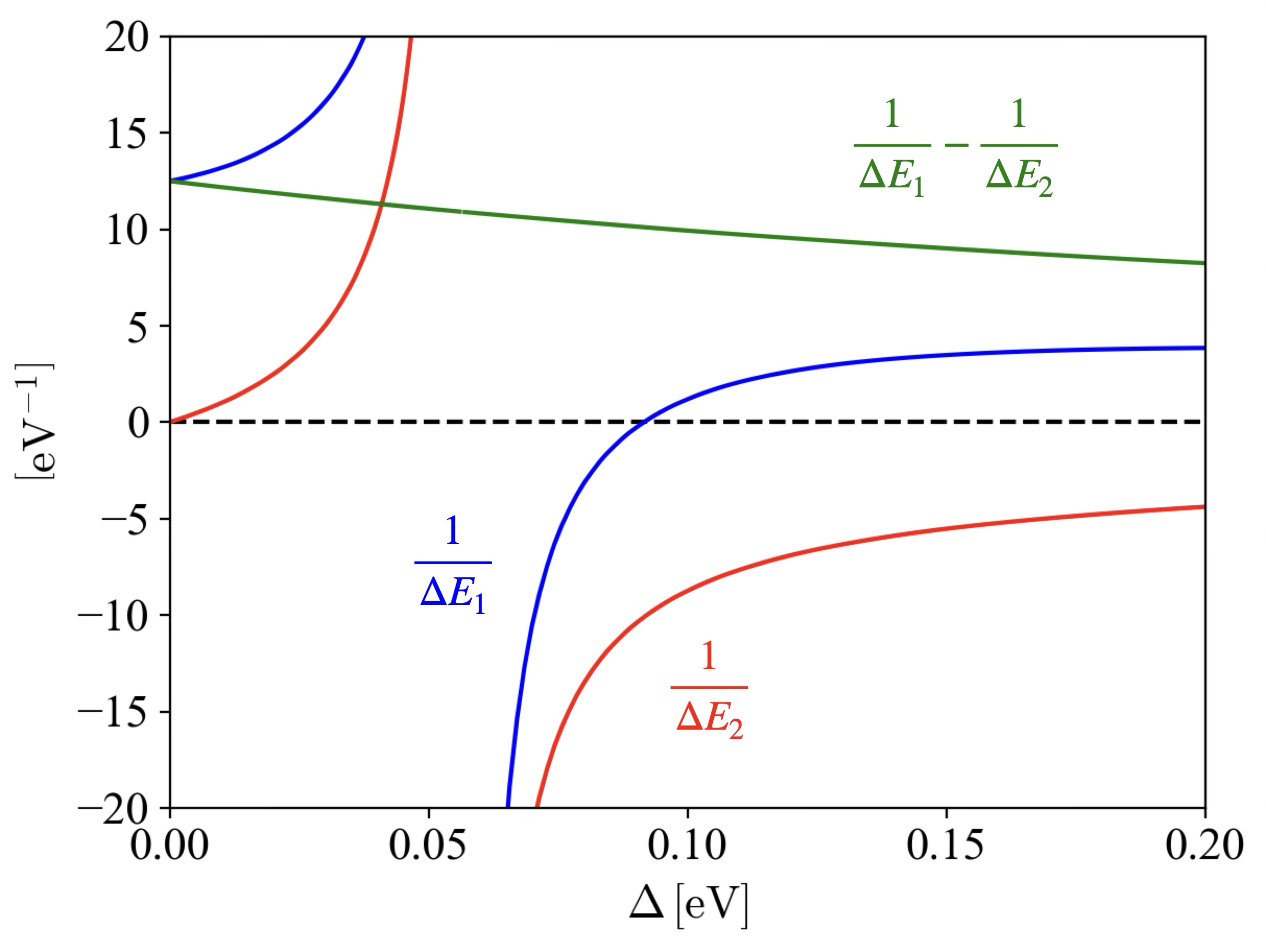}
    \caption{Inverse energy terms from perturbation theory, as defined in Eq. (\ref{deltaE}). We use $\epsilon = 0.12$ eV, $s_1=-1.62$ and $s_2=0.5.$}
    \label{fig:inversenergy}
\end{figure}

Microscopic details of the tight-binding model are inherited by the effective model through $\lambda,$ and are responsible for determining which of the $\Phi=\pi$ phases are stabilized. Indeed, at $\lambda=0$ the $\Phi = \pi$ phases CBO$-$, LCBO$+,$ and NLCBO are degenerate. The degeneracy is broken by $\lambda,$ favouring NLCBO within a small range of chemical potentials in between vH1 and vH2. In this regime, the two lowest-energy bands are deep within the Fermi sea. Of interest therefore is the level nearest to the Fermi energy, in between vH1 and vH2. Performing perturbation theory on $\lambda$, subject to the total-phase constraint $\phi_1+\phi_2+\phi_3=\pi,$ we find the second-order correction to this band to be
\begin{align}
    & \delta E = \sum_{\mbfk_{\rm occ}}\frac{\lambda^2}{9}\Bigg( \left[\frac{1}{\Delta E_1}\right](k_1^2+k_2^2+k_3^2) \\
   &+  \left[\frac{1}{\Delta E_2}\right]2\Big(k_1k_2\cos(2\phi_1) + k_2k_3\cos(2\phi_2) + k_3k_1\cos(2\phi_3) \Big) \Bigg), \nonumber
\end{align}
where $\delta E \equiv E (\lambda) - E (\lambda=0)$. The $\phi_i$ can be chosen in many ways to satisfy the total-phase condition. For CBO$-,$ $\phi_1=\phi_2=\phi_3=\pi$, i.e. $(-\Delta, -\Delta, -\Delta)$. For LCBO$+,$ $\phi_1=\phi_2 =\phi_3 = \pi/3.$ For NLCBO, we may choose $\phi_1=0$ and $\phi_2=\phi_3=\pi/2,$ breaking the rotational symmetry. We thus obtain:
\begin{align}\label{delta_Energy}
    \delta E_{\rm CBO-} & = \frac{\lambda^2}{6}\sum_{\mbfk_{\rm occ}} \left[\frac{1}{\Delta E_1}-\frac{1}{\Delta E_2}\right] (k_x^2+k_y^2),\nonumber \\
    \delta E_{\rm LCBO+}
        & = \frac{\lambda^2}{6} \sum_{\mbfk_{\rm occ}} \left( \left[ \frac{1}{\Delta E_1}-\frac{1}{\Delta E_2} \right](k_x^2+k_y^2) + \frac{3}{2}\frac{k_x^2+k_y^2}{\Delta E_2} \right), \nonumber \\
    \delta E_{\rm NLCBO}
        & = \frac{\lambda^2}{6} \sum_{\mbfk_{\rm occ}} \left( \left[\frac{1}{\Delta E_1} - \frac{1}{\Delta E_2}\right]\left(k_x^2+k_y^2\right) + \frac{8}{3}\frac{k_x^2}{\Delta E_2} \right),
\end{align}
where
\begin{align}\label{deltaE}
    \left[\frac{1}{\Delta E_1}\right] & \equiv \left[ \frac{1}{2\Delta(s_1-s_2)+2\epsilon} -\frac{2}{\Delta(s_1+2s_2)-2\epsilon} \right], \nonumber \\
    \left[\frac{1}{\Delta E_2}\right] & \equiv \left[ \frac{1}{2\Delta(s_1-s_2)+2\epsilon} +\frac{1}{\Delta(s_1+2s_2)-2\epsilon} \right].
\end{align}
Provided $\Delta$ is sufficiently large, we find $\tfrac{1}{\Delta E_1}>0$ and $\tfrac{1}{\Delta E_2}<0$ (see Fig. \ref{fig:inversenergy}). Written this way, we see that each order acquires a concave up parabolic and isotropic component of the dispersion $ \tfrac{\lambda^2}{6}\left[\tfrac{1}{\Delta E_1}-\tfrac{1}{\Delta E_2}\right](k_x^2+k_y^2)>0.$ This is the only contribution to the CBO$-$ band at second order in $\lambda.$ The LCBO$+$ phase acquires an additional isotropic component which suppresses the band curvature near $k=0$ since $\tfrac{1}{\Delta E_2}<0.$ The nematic phase spontaneously breaks the isotropy, developing a strong anomalous dispersion along $k_x.$ Because $\tfrac{1}{\Delta E_2}<0,$ the band associated with NLCBO disperses downwards more strongly than the competing phases. Under certain conditions, this anomalous dispersion enables NLCBO to stabilize over CBO$-$ or LCBO$+$. This mechanism relies on three factors: i) sufficiently large $\lambda$ to control the strength of the dispersion; ii) the sign of $\tfrac{1}{\Delta E_2}<0$ to lower the band energy; and iii) a chemical potential that partially fills the band (if the band is fully occupied, one can show that LCBO$+$ has the lowest free energy). The fine-tuning required to simultaneously satisfy these three conditions is likely responsible for the small phase space in which NLCBO develops.

The discussion presented in this section is limited to terms of order $O(\lambda^2)$, while higher order contributions are neglected. Thus it should be viewed only as providing intuition for the tendency of the system to develop the nematic order. In addition, we have assumed an equal magnitude of $\Delta$ for the three competing orders to isolate the role of the complex phases. Allowing for inequivalent bond amplitudes, NLCBO typically stabilizes with $\Delta’ > \Delta.$ Although the phase optimization is the primary factor in stabilizing nematic order, this amplitude asymmetry further enables NLCBO to become energetically favourable over the competing phases.

Our analysis demonstrates the frustration inherent in the system through the various possible choices for $\phi_i.$ The phase frustration that drives nematicity in the patch model persists when additional bands are included and the calculation is extended over the full Brillouin zone (see  Sec. \ref{sec:9band}). A free energy accounting for both the amplitude and phase of the order parameter is obtained through symmetry considerations in Ref. \cite{Denner2022}, and also suggests nematic order arising from phase frustration. Our results are obtained microscopically, complementing this symmetry-derived free energy.

\section{Application to Vanadium Kagome metals:
Tight-binding model and order parameters}
\label{sec:9band}

\begin{figure}
    \centering
    \includegraphics[width=\linewidth]{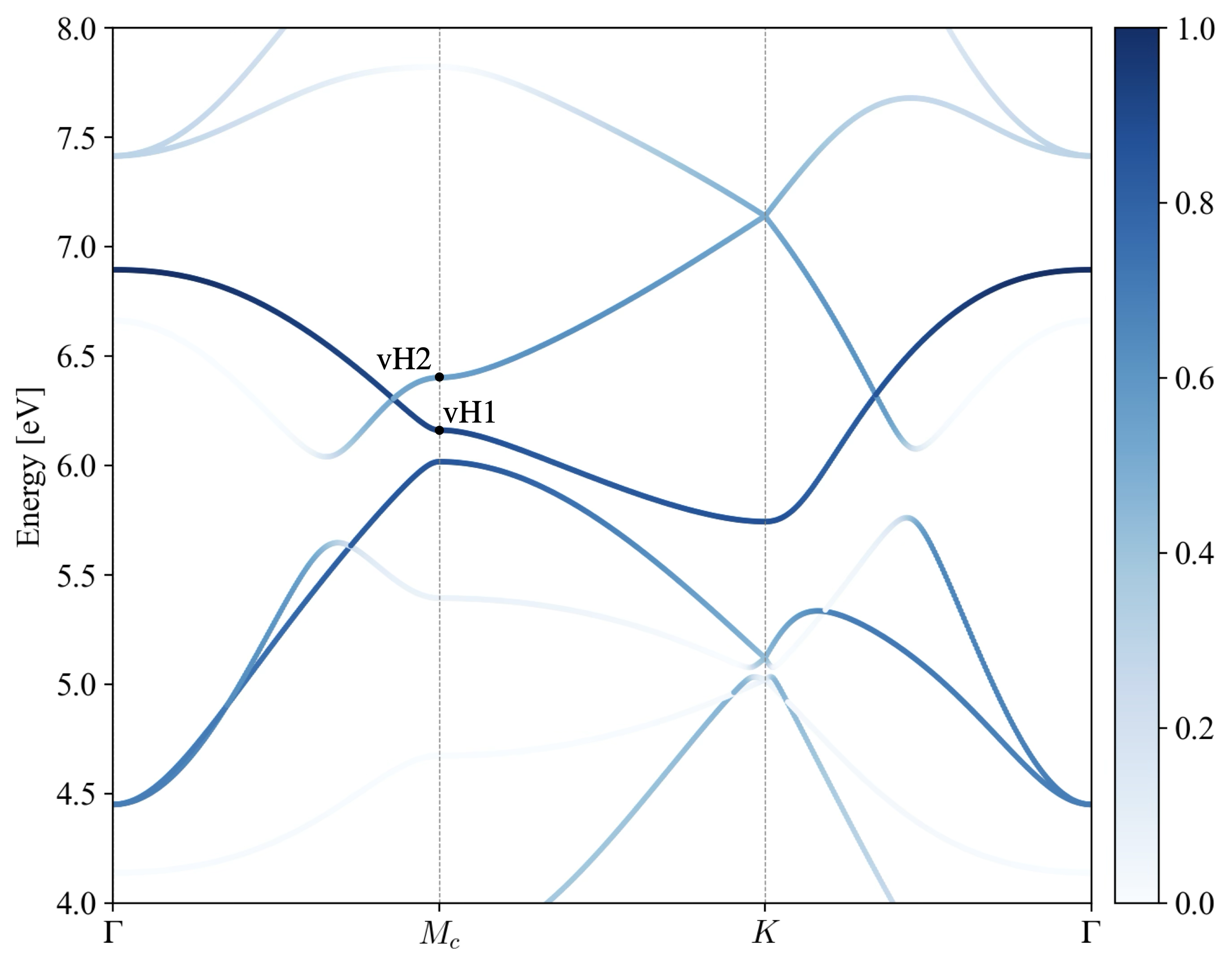}
    \caption{Band structure of the nine-band tight-binding model. Colour denotes the $d$-orbital weight.}
    \label{fig:bandstructure}
\end{figure}

To demonstrate the robustness of our analysis within the effective model, we now turn to a tight-binding description defined over the full Brillouin zone. The tight-binding model, described in detail in Ref. \cite{HeqiuPRB}, accounts for five $d$ orbitals at each of the three vanadium sites, plus three $p$ orbitals at the five antimony sites within the unit cell, i.e. 30 bands in total. The hopping amplitudes and on-site potentials are obtained through DFT. The 30-band model is then reduced to a more manageable nine-band model by keeping only the dominant orbitals near the $M$ points $\mathbf{M}_A=(-\pi/a,-\pi/\sqrt{3}a),$ $\mathbf{M}_B=(\pi/a,-\pi/\sqrt{3}a),$ $\mathbf{M}_C=(0,2\pi/\sqrt{3}a).$ The resulting model contains one $d$ orbital per vanadium site and three $p$ orbitals at each of the two out-of-plane antimony sites. The band structure of the nine-band model is plotted along the path $\Gamma\rightarrow M\rightarrow K \rightarrow\Gamma$ in Fig. \ref{fig:bandstructure}.

The nine-band model captures one $p$-type (vH1) and one $m$-type (vH2) van Hove singularity at each $M$ point. The dominant $d$ orbital contribution to the vH1 wavefunction at $M_\alpha$ is denoted $\tilde{d}_\alpha,$ and consists of a particular linear combination of the atomic $d_{xz}$ and $d_{yz}$ orbitals. The dominant $p$ orbital contributions to the vH1 wavefunction at $M_\alpha$ are denoted $\tilde{p}_{\sigma\alpha},$ where $\sigma=1,2$ denotes the antimony sublattice. The $\tilde{p}_\alpha$ orbitals consist of linear combinations of the atomic $p_x$, $p_y,$ and $p_z$ orbitals above and below the Kagome plane with mirror $M_z$ eigenvalue $-1$ \cite{HeqiuPRL}.

The $2a\times 2a$ CDW order is induced by a NN interaction between the vanadium ${d}$ orbitals, 
\begin{equation}
    H_V = V\sum_{\alpha\neq\beta}\sum_{\langle \mathbf{R}\mathbf{R'} \rangle} d_{\mathbf{R},\alpha}^\dagger d_{\mathbf{R},\alpha} d_{\mathbf{R'},\beta}^\dagger d_{\mathbf{R'},\beta}.
\end{equation}
Here, $\mathbf{R}$ and $\mathbf{R'}$ label neighbouring unit cells, $\alpha,\beta=A,B,C$ denote the Kagome sublattices, and $V$ is the NN interaction strength. The interaction can be mean-field decoupled as \cite{HeqiuPRB}
\begin{eqnarray}
    H_V^{\rm MF}&=& -\sum_{\mathbf k} \left(\Delta_{\alpha\beta}(1-e^{i\mathbf k \cdot \mathbf d_{\alpha\beta}})d^\dagger_{\mathbf k-\mathbf Q_{\alpha\beta},\beta}d_{\mathbf k,\alpha}+h.c. \right)\nonumber \\
    &&+2N_c \frac{|\Delta_{\alpha\beta}|^2}{V} ,\label{Hvmf} 
\end{eqnarray}
with $\Delta_{\alpha\beta}$ according to Eq. (\ref{Deldef}).

\subsection{Phase diagram}
\label{phaseD9x9}

\begin{figure}
    \centering
    \includegraphics[width=\linewidth]{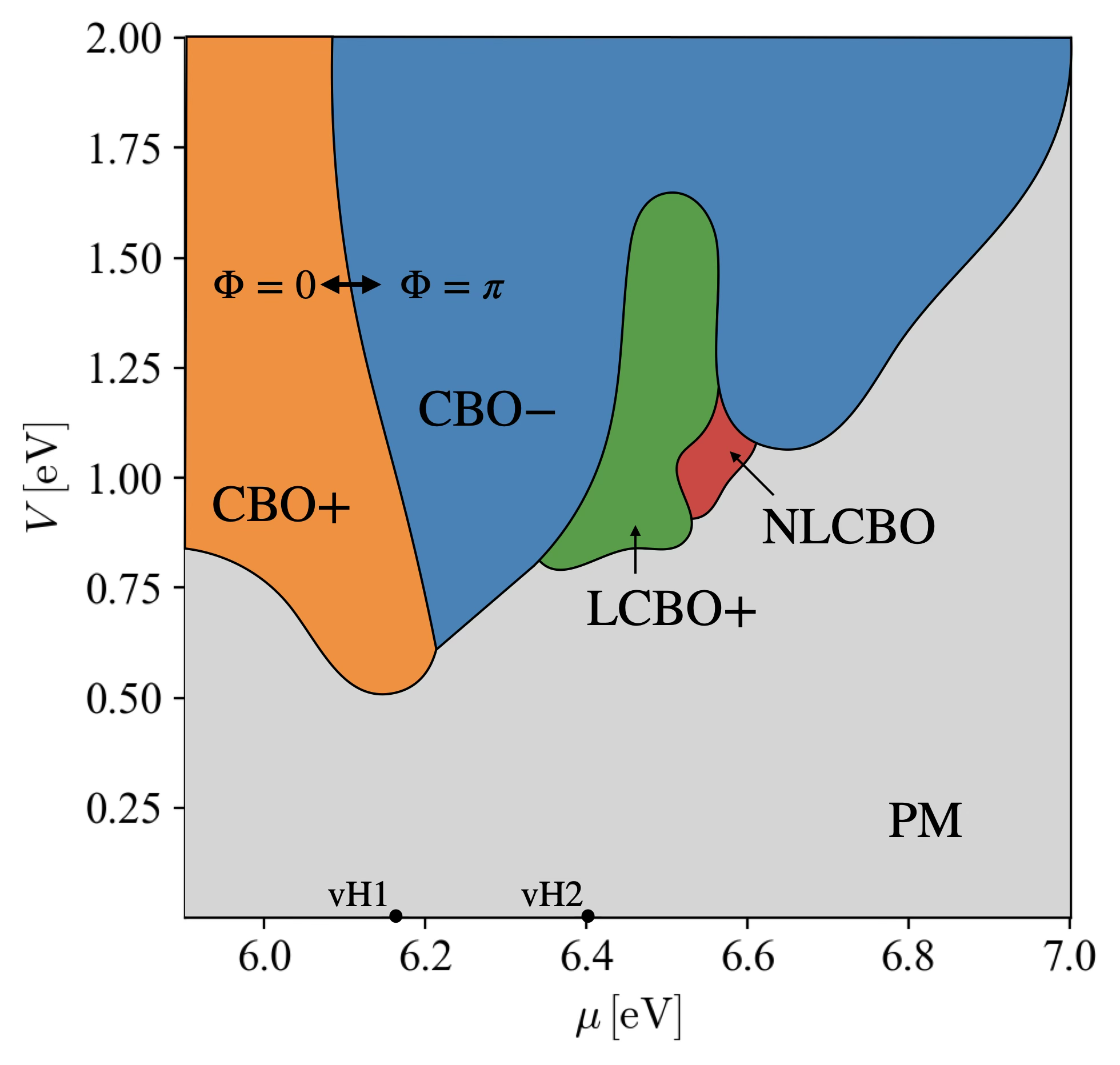}
    \caption{Phase diagram of the tight-binding model at $T=90$ K. NLCBO emerges in a small pocket above vH2. The total phase transitions from $\Phi=0$ to $\Phi = \pi$ across $\mu\approx \rm{vH1}$.}
    \label{fig:9x9a}
\end{figure}

In Fig. \ref{fig:9x9a}, we present the mean-field phase diagram of the tight-binding model augmented by the interaction term $H_V^{\rm MF}.$ The phase diagram is plotted as a function of the chemical potential $\mu$ and the interaction strength $V,$ at a finite temperature of $T = 90$ K to smooth out artifacts of the $k$-space discretization. Many of the key features of the phase diagram obtained from the effective model persist in the tight-binding model. Unlike in the effective model, where we are free to tune $\lambda,$ the band curvature near $M$ is predetermined ($\lambda\approx 0.35\,{\rm{eV\cdot}}a$). The primary difference between the two models lies in the region of phase space where NLCBO appears. This discrepancy likely arises from the more intricate band structure of the tight-binding model. In particular, the delicate energy balance necessary to stabilize nematic order described in Sec. \ref{FSFX} may be disrupted. Compared to the effective theory, the tight-binding model exhibits ordered phases over a wider range of chemical potentials due to the presence of these additional bands. Accordingly, the interaction strength is renormalized in the effective theory because of the different bandwidths, and therefore their values in the two models cannot be directly compared.

\begin{figure}
    \centering
    \includegraphics[width=0.9\linewidth]{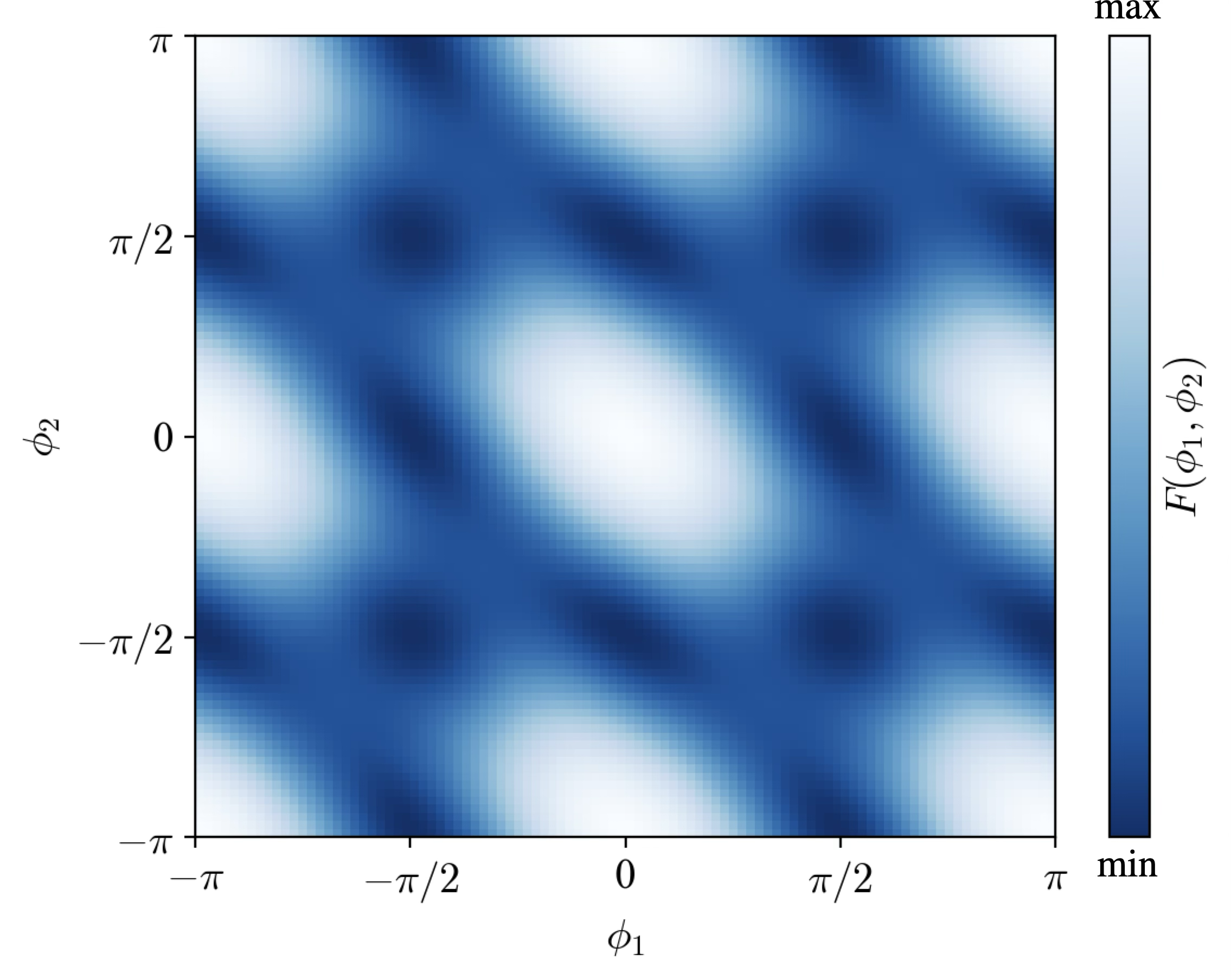}
    \caption{Free energy landscape for $\Phi=\pi$ phases at $V=1.0$ eV and $\mu=6.55$ eV.}
    \label{fig:freeNRGlanscape}
\end{figure}

To justify our identification of the nematic phase as possessing the particular phase structure $(\Delta,i\Delta',i\Delta'),$ we examine a point in phase space within the red region labeled NLCBO in Fig. \ref{fig:9x9a}. Assuming equal $\Delta=\Delta'$ to isolate the phase structure, we parameterize the order parameter by $\Delta(e^{i\phi_1},e^{i\phi_2},e^{i\phi_3}).$ Imposing the constraint $\Phi = \phi_1+\phi_2+\phi_3=\pi,$ we examine the free-energy landscape of the tight-binding model as a function of $\phi_1$ and $\phi_2$ in Fig. \ref{fig:freeNRGlanscape}. We find that the free-energy minima are found at permutations of $(\phi_1,\phi_2,\phi_3)=(0,\frac{\pi}{2},\frac{\pi}{2}),$ corroborating our identification of the $\phi_i$ in NLCBO.

We note that there are few other nematic phases that emerge within our calculations, mostly at lower temperatures and near phase boundaries. These phases are characterized by purely real order parameters of the form $(\Delta,\Delta',\Delta'),$ or $(\Delta,0,0).$ However, their corresponding phase space is very small (see Appendix \ref{Numerical}). Calculations were performed on a $30\times30$ momentum-space grid, which we found sufficiently dense to achieve convergence.

\section{Discussion and Outlook}
\label{sec:Conclusion}

In this work, we have investigated the tendency of the Kagome metals to develop $2a\times 2a$ nematic loop-current bond order. The nematic phase is characterized by coexisting charge-bond order along one of the three Kagome bonds and loop-current order along the other two, thereby breaking two of three mirror planes and time-reversal symmetry. To explain the emergence of this nematic order, we analyzed an effective patch model that includes two van Hove singularities at each of the three inequivalent $M$ points. The effective model reveals the frustration associated with the relative phases of the three-component ($3\mathbf{Q}$) order parameter. Within this framework, the free energy is minimized by the development of an elongated Fermi surface that lowers the sixfold symmetry to twofold, giving rise to nematic order that may be observable in angle-resolved photoemission spectroscopy (ARPES). To demonstrate the robustness of the effective model, we employ a nine-band tight-binding model with NN interactions, finding that the nematic order survives in a narrow region of phase space within mean-field theory.

Looking ahead, this study motivates several directions for future research. Our results indicate that NLCBO emerges only within a narrow region of phase space where the energetics become favourable. A key sensitivity of our theory lies in the energy spacing $2\epsilon$ between the two van Hove singularities, which governs the strength of the anomalous dispersion in the nematic phase, together with the phase imbalance among the three complex order parameters. This suggests that pressure, strain, and doping studies could tune this spacing \cite{LaBollita2021,Sim2024}, providing a potential route to access the phase transition into NLCBO. Furthermore, the coupling between the two van Hove singularities, $\lambda$, which serves as a tuning knob for the emergence of nematic order, may also be adjustable under pressure.

Within the $\rm{A}\rm{V}_3\rm{Sb}_5$ family, $s_1$ and $s_2$ depend on the choice of the alkali atom ($\rm{A}=\rm{K,\,Rb,\,Cs}),$ which will be relevant for satisfying the energetics necessary to stabilize NLCBO. The signs of $s_1<0$ and $s_2>0,$ however, are determined by the mirror symmetries of vH1 and vH2. Relaxing this constraint, the effective model may be applicable to other Kagome metals possessing van Hove singularities with different mirror symmetries. Interestingly, the isostructural titanium-based Kagome metal $\rm{Cs}\rm{Ti}_3\rm{Bi}_5$ shows nematicity in the superconducting state, which does not evolve from a parent CDW \cite{YangH2024}. The van Hove singularities in $\rm{Cs}\rm{Ti}_3\rm{Bi}_5$ are located further away from the Fermi level than in $\rm{A}\rm{V}_3\rm{Sb}_5,$ pointing to an entirely different mechanism for nematic order.

The nature of superconductivity in Kagome metals remains an open research question. Of particular interest are the unusual pressure-dependent properties, including nematicity and the double-dome dependence of the superconducting transition temperature $T_c$ \cite{YuFH2021,ZhengL2022,ZhouReview2024}. Since the superconducting state likely inherits key characteristics of its parent CDW upon cooling, further investigation into the nature of the parent CDW is essential. Although we leave to future work the study of the superconducting state that may emerge from the proposed nematic order, the rich pressure-dependent phase diagrams suggest a strong sensitivity of the Kagome metals to microscopic parameters, underscoring the need for continued theoretical and experimental exploration.

\textit{Acknowledgments.} This work is supported by the NSERC Discovery Grant No. 2022-04601. H. Y. K. acknowledges support from the Canada Research Chairs Program No. CRC-2019-00147. This research was enabled in part by support provided by Compute Ontario, Calcul Québec, and the Digital Research Alliance of Canada.

\appendix

\section{$\Phi=0$ to $\Phi=\pi$ transition}\label{Phitransition}

The unperturbed eigenvalues for $\Phi=0$ and $\Phi=\pi$ phases can be written
\begin{align}
    E^{(1)}_0 & = -\epsilon -\mu \pm2s_1|\Delta| , & E_{1,2}^{(1)} = -\epsilon -\mu  \mp s_1|\Delta|  \nonumber\\
    E^{(2)}_0 & = \epsilon -\mu  \pm 2s_2|\Delta| , & E_{1,2}^{(2)}  = \epsilon -\mu \mp s_2|\Delta| ,
\end{align}
where the upper (lower) signs correspond to $\Phi=0$ ($\Phi=\pi$). Upon tuning $\mu$ to vH1, developing finite $\Delta$ drives the vH1-derived bands away from the Fermi level. Whether the degenerate level or the non-degenerate level is driven above or below depends on $\Phi$. Since $s_1<0,$ if $\Phi=0,$ the non-degenerate level $E_0^{(1)}$ is driven below the FS and the internal energy of the system is (to zeroth order) $U_0=-\epsilon-\mu-2|s_1||\Delta|.$ If $\Phi=\pi,$ the degenerate level $E_{1,2}^{(1)}$ is instead driven below the FS and the internal energy of the system is $U_\pi=2\left(-\epsilon -\mu -|s_1||\Delta|\right).$ $\Phi=0$ is favoured over $\Phi=\pi$ if $U_0<U_\pi,$ which simplifies to $\mu<-\epsilon.$ Thus, there is a phase transition from $\Phi=0$ to $\Phi=\pi$ as $\mu$ passes through vH1.

The sign of $s_1<0$ is crucial in determining whether $\Phi=0$ or $\Phi=\pi$ is favoured for $\mu$ below vH1. 
If $s_1$ were positive, the transition would be reversed. That is precisely the situation at vH2, where $s_2>0.$ In this case, we expect $\Phi=\pi$ to be favoured for $\mu<\epsilon$ and $\Phi=0$ to be favoured for $\mu>\epsilon.$ The relative sign between $s_1$ and $s_2,$ which originates from the symmetry-imposed form of the wavefunction at $M,$ conspires to select $\Phi=\pi$ for $\mu$ in between vH1 and vH2 and $\Phi=0$ for $\mu$ outside. The sharp transition from $\Phi=0$ to $\Phi=\pi$ in the effective model is softened in the tight-binding model due to the presence of additional bands.

\section{Numerical details}\label{Numerical}

In principle, the free energy is to be minimized over a six-dimensional space, corresponding to the real and imaginary components of the three order parameters $\Delta_{\alpha\beta}.$ The number of components to minimize can be reduced by assuming a particular symmetry, e.g. for real CBO with $\Delta_{AB}=\Delta_{BC}=\Delta_{CA}=|\Delta|$ the minimization is one-dimensional. We seek solutions corresponding to the phases summarized in Fig. \ref{fig_NLCBO}. When more than one solution is found, we select the solution with the lowest free energy. Mean-field terms proportional to the total density simply renormalize the chemical potential have been neglected, which only lead to shifts in the phase boundaries of our phase diagrams.

For the effective model, the dispersion of the bands about the $M$ points is controlled by $\lambda$. The limiting case $\lambda\rightarrow 0$ corresponds to a set of CDW bands with no dispersion. As $\lambda$ is increased, the CDW bands acquire a dispersion. As we have demonstrated, nonzero $\lambda$ is required to stabilize NLCBO. The ``true'' value of $\lambda$ consistent with the bands of the $9\times 9$ tight-binding model gives $\lambda\approx 0.35 \,{\rm{eV\cdot}}a.$ We note that only the magnitude of $\lambda$ matters --- its complex phase is unimportant.

There is a subtlety in the choice of $k_{\rm cut}.$ From Eq. (\ref{explicit}), we see that the momentum is always scaled by $\lambda.$ Therefore, it is only the product $\lambda k$ (which has units of energy) that is important, and tuning the momentum cutoff is equivalent to tuning $\lambda.$ For this reason, we fix $k_{\rm cut} = 1$ and tune $\lambda$ within the effective model. The mean-field calculations converge sufficiently with a $k$-mesh of $\sim 125$ points, i.e. phase boundaries do not change as the grid density is increased. In the main text, we present the results for a $k$-space sampling of $\sim500$ points.

\begin{figure}
    \centering
    \includegraphics[width=\linewidth]{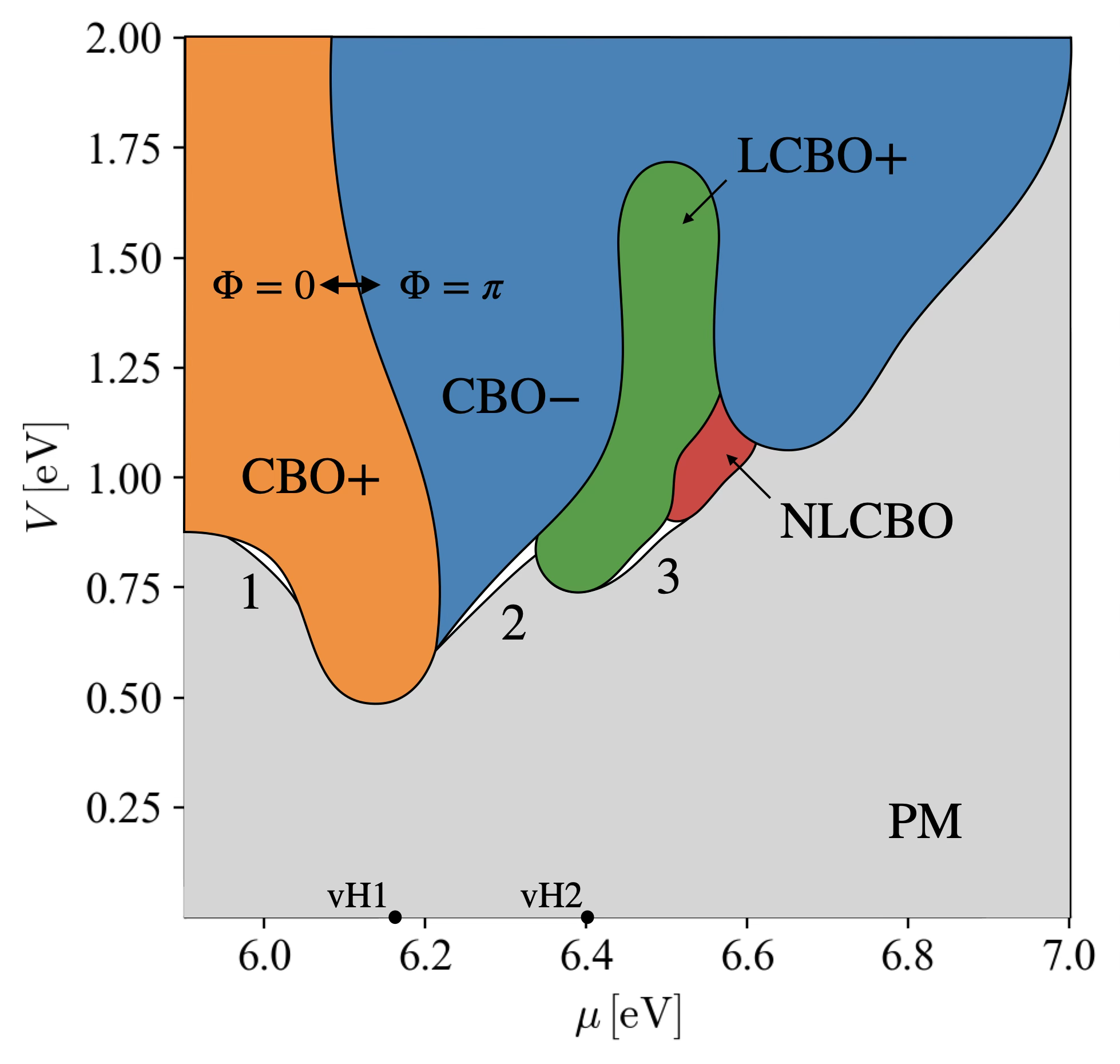}
    \caption{Phase diagram of the tight-binding model at $T=0$ K. The additional nematic phases are characterized by order parameters of the form 1: $(\Delta,\Delta',\Delta'), \Delta\gg\Delta' \in \mathbb{R}$ and CBO$+$ sign structure, 2: $(\Delta,\Delta',\Delta'), \Delta\gg\Delta' \in \mathbb{R}$ and CBO$-$ sign structure, and 3: $(i\Delta,0,0), \Delta\in \mathbb{R}.$}
    \label{fig:9x9a_OtherPhases}
\end{figure}

For the tight-binding calculation, the non-interacting Hamiltonian is a $9\times 9$ matrix, corresponding to three vanadium $d$ orbitals and six antimony $p$ orbitals. The ordering induced by the wavevectors $\mathbf{Q}=\mathbf{M}_A,\mathbf{M}_B,\mathbf{M}_C$ couples states at $\mbfk$ to three other states at $\mbfk +\mathbf{Q},$ resulting in a $36\times36$ matrix to be minimized over the reduced Brillouin zone. In Sec. \ref{phaseD9x9}, we mentioned that other nematic phases appear as mean-field solutions at lower temperatures. We find that these phases do persist upon increasing the $k$-mesh density, but their corresponding phase space remains extremely narrow. For this reason we do not believe these phases to be particularly robust beyond the mean-field approximation. We plot the phase diagram of the tight-binding model at zero temperature in Fig. \ref{fig:9x9a_OtherPhases}. In comparison to the $T=90$ K calculation of Fig. \ref{fig:9x9a}, the phase boundaries are slightly modified. The additional nematic phases (marked in white) appear at phase boundaries.

\bibliography{biblio}
\end{document}